# An adsorbed gas estimation model for shale gas reservoirs via statistical learning

Yuntian Chen,[a,1] Su Jiang,[b,1] Dongxiao Zhang,[a,*] and Chaoyang Liu[c]

[a] ERE and BIC-ESAT, College of Engineering, Peking University, No. 5, Yiheyuan Road, Beijing 1000871, China

[b] Department of Energy Resources Engineering, Stanford University, Stanford, California 95305, U.S.A.

[c] Department of Materials Science and Engineering, Massachusetts Institute of Technology, Cambridge, Massachusetts 02139, U.S.A.

[*] Corresponding author at: College of Engineering, Peking University, No. 5, Yiheyuan Road, Beijing 1000871, China. Tel.: +86 10 6275-7432; fax: +86 10 6275-6607.

E-mail address: dxz@pku.edu.cn (D. Zhang)

[1] Authors have made equal contributions to this work.

**Graphical Abstract:**

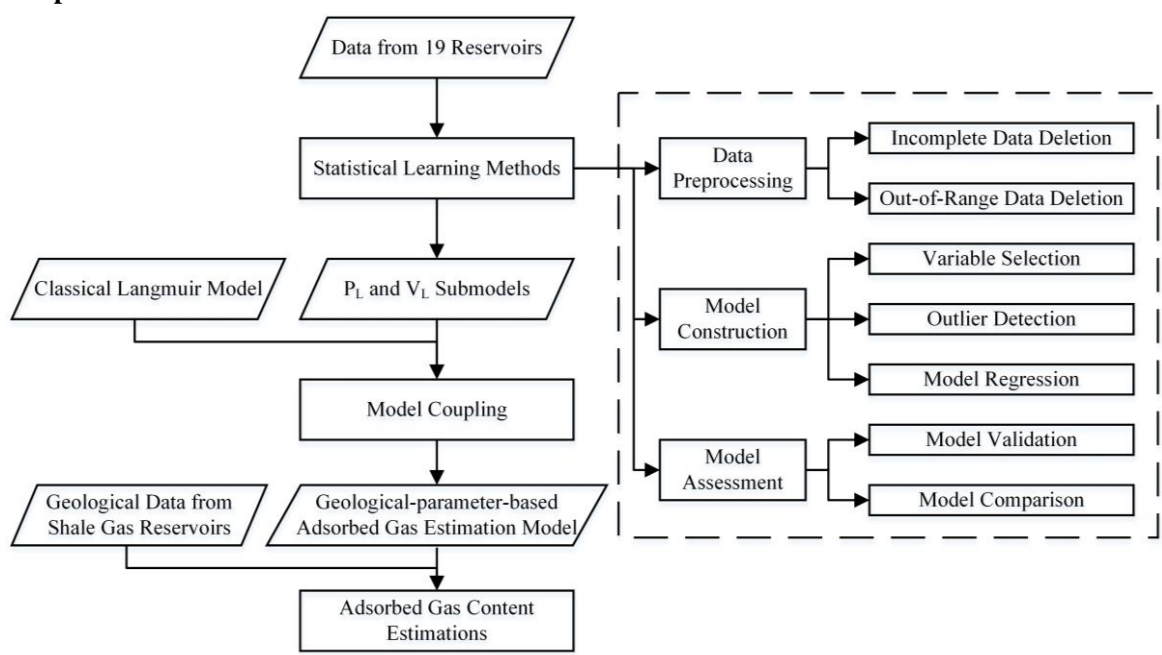

**Abstract:**

Shale gas plays an important role in reducing pollution and adjusting the structure of world energy. Gas content estimation is particularly significant in shale gas resource evaluation. There exist various estimation methods, such as first principle methods and empirical models. However, resource evaluation presents many challenges, especially the insufficient accuracy of existing models and the high cost resulting from time-consuming adsorption experiments. In this research, a low-cost and high-accuracy model based on geological parameters is constructed through statistical learning methods to estimate adsorbed shale gas content. The new model consists of two components, which are used to estimate Langmuir pressure ($P_L$) and Langmuir volume ($V_L$) based on their quantitative relationships with geological parameters. To increase the accuracy of the model, a "big data" set that consists of 301 data entries was compiled and utilized. Data outliers were detected by the K-Nearest Neighbor (K-NN) algorithm, and the model performance was evaluated by the leave-one-out algorithm. The proposed model was compared with four existing models. The results show that the novel model has better estimation accuracy than the previous ones. Furthermore, because all variables in the new model are not dependent on any time-consuming experimental methods, the new model has low cost and is highly efficient for approximate overall estimation of shale gas reservoirs. Finally, the proposed model was employed to estimate adsorbed gas content for nine shale gas reservoirs in China, Germany, and the U.S.A.

## 1. Introduction

Shale gas, or natural gas trapped within shale formations, is a type of relatively clean energy resource. It consists mainly of methane and burns cleaner than other kinds of hydrocarbon fuel. As an alternative fuel, shale gas is attracting increased attention globally [1-8]. Shale gas consists of free gas, adsorbed gas, and solution gas: free gas is the shale gas in the pore space within the shale rock; adsorbed gas is a significant quantity of gas adsorbed on the surface of organics and clays in the shale formation; and solution gas is the gas dissolved in the reservoir water or oil. The volume of solution gas is governed by pressure and temperature. As the pressure drops below the bubble point, the gas dissolved in the liquid begins to be released and becomes free gas. **Fig. 1** illustrates these three kinds of shale gas. In recent years, the production of shale gas has increased significantly. For instance, in 2000, shale gas only occupied 1.6% of gas production in the U.S.A. [9], while this percentage rose to 47% in 2015 [10]. In China, the total production of shale gas was only 200 million m³ in 2013. However, it increased to 1.3 billion m³ in 2014 [11] and 4.47 billion m³ in 2015 [12]. Shale gas is thus expected to play an even more critical role in the world's energy supply in the near future.

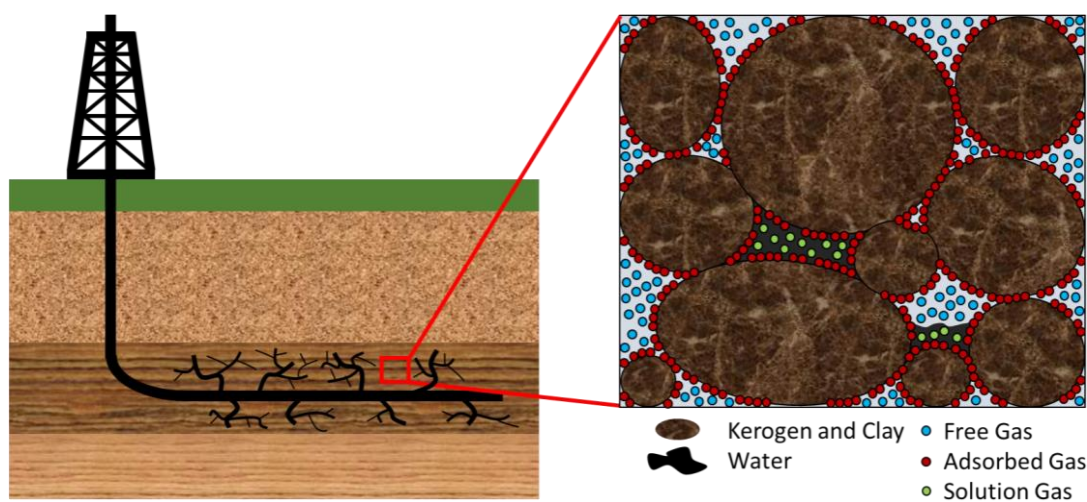

**Fig. 1.** Illustration of free gas, adsorbed gas, and solution gas in shale formations.

There are many challenges involved in shale gas development and evaluation processes. One of the most important problems is to provide an accurate resource estimation. For instance, the U.S. Energy Information Administration (EIA) and the Chinese Ministry of Land and Resources have quite different evaluations of China shale gas resources, which are 31.6 trillion m³ and 25.1 trillion m³, respectively [13, 14]. The uncertainty regarding resource estimation will not only affect the gas well site selection on a micro level, but will also influence the national and industrial policy-making process on a macro level. Thus, it is crucial to find a way to accurately evaluate shale gas resources around the world. This research emphasizes adsorbed gas because it accounts for 20% to 80% of the total gas [15-17]. Adsorbed gas also includes solution gas, which is only a small portion of the total gas content [14, 18]. Regarding free gas, many researchers have already proposed accurate estimation models [18-20].

In gas content estimation, first principle methods and empirical models are common options. The first principle methods are still not thoroughly developed, because of the complexity of storage mechanisms and the lack of understanding of the influence of numerous factors, such as thermal maturity ($R_o$) and reservoir temperature (T). Thus, it is challenging to build an accurate theoretical model. Regarding the empirical models, the Langmuir model (**Eq. (1)**) is the most commonly used [21, 22]. Its primary advantage is that, once Langmuir pressure ($P_L$) and Langmuir volume ($V_L$) are determined, the Langmuir adsorption isothermal curve is ascertained, which makes it easy to calculate adsorbed gas content under any reservoir pressure. Nevertheless, adsorption experiments are probably the only effective methods to obtain Langmuir parameters. However, these experiments are very time-consuming, as the adsorption process within the microscale and nanoscale shale pores is slow [23], and the coring process to obtain the experiment sample is expensive. These challenges cause difficulties in determining the corresponding Langmuir parameters and high uncertainties regarding adsorbed gas content evaluation. In addition, most of these experiments are complicated and are subject to experimental errors. For example, leakage always occurs in the coring process, which affects the accuracy of the adsorption experiments. Because both the first principle methods and empirical models have difficulties in gas content estimation, the salient question is: are there any alternatives that are reasonably accurate, but do not rely on any site-specific adsorption experiments?

$$V = \frac{P \cdot V_L}{P + P_L} \tag{1}$$

In this work, we proposed to use widely available geological parameters to estimate resource volume via statistical learning. After years of development, statistical learning has become a powerful tool to build models [24-36], but it has never been applied to shale gas resource evaluation. Statistical learning is a commonly used method for Big Data Analytics, and it is effective in finding a predictive function based on a big data set. Statistical learning has many successful applications in various fields, such as computer vision, sales forecasting, and bioinformatics. It is not only utilized to uncover hidden patterns and unknown correlations between variables, but is also utilized to alleviate the influence of data uncertainty resulting from the operation process. For the problem that we are considering, geological parameters, such as total organic carbon (TOC), thermal maturity ($R_o$, vitrinite reflectance) and reservoir temperature (T), are much easier to obtain compared with Langmuir parameters. Thus, one promising strategy is to use these easily-obtained geological parameters to estimate the Langmuir parameters with statistical learning methods.

Attempts have been made to discern the qualitative correlation between geological parameters and Langmuir parameters. The positive correlation between $V_L$ and TOC has been shown in the literature [37-40], and a similar positive correlation between $R_o$ and $V_L$ is found [23]. The qualitative influences of TOC, $R_o$, and T on Langmuir parameters are revealed by Zhao et al. [41]. Regarding $P_L$, Kong et al. identified the effect of T, $R_o$ and porosity on the shale gas adsorption process, and discussed the determining physical mechanisms [42]. Hao et al. studied the influence of T on $P_L$, as well [43]. Although the qualitative effect of different factors has been widely studied, the quantitative relationships remain unclear. Attempts have been made to use linear regression to estimate $V_L$ with TOC as the only independent variable [44-46]. Although TOC is essential in adsorption, previous studies have shown that TOC is not the only influencing factor [23, 41]. The dataset of these works are not abundant as well. For example, there are only 7-10 data points from shale gas reservoirs in these studies. As for $P_L$, Xia et al. attempted to build a first principle model based on isoteric heat, standard entropy, and temperature [47]. Nevertheless, the variables used in this theoretical model are difficult to measure, which restricts model applicability. Later, Zhang et al. utilized T as an independent variable to estimate $P_L$ [46]. T is a significant factor since adsorption is an exothermic process. However, it is not comprehensive to consider T as the only independent variable. In addition, this model's dataset only has six data points.

In conclusion, many studies in the last decade estimated Langmuir parameters by using geological parameters, but these quantitative models usually present three unresolved problems. The first one is inadequate sample size. Owing to the high cost of adsorption experiments, the datasets of previous studies are often too small to obtain a reliable and stable model. Some of these data are from coal bed methane and are not suitable for shale gas analysis [44, 46]. The second one concerns factor selection. Numerous previous studies consider factors non-comprehensively [44-46]. Specifically, most of these studies only focused on TOC. This is partially reasonable since TOC constitutes the main hydrocarbon-generation matter and gas carrier [48], but TOC alone cannot describe Langmuir parameters sufficiently, and other factors should be considered [23]. The last problem is the lack of general applicability. Most of the previous work only has data that cover a certain reservoir or even a single well, and the validation test is always overlooked [44-47]. In fact, among all of the three problems, inadequate sample size presents the biggest obstacle, as it partially causes the other two problems.

The main objective of this study is to solve the three problems above and build an adsorbed gas content estimation model based on geological parameters via statistical learning. The model-building process is shown in **Fig. 2**. To solve the inadequate sample size problem, 301 adsorption experiment data sets were compiled and analyzed from 19 different reservoirs in seven countries. Our data are much bigger than those in previous studies, which usually have less than 10 data sets. Once the sample size is enlarged, statistical learning may become more effective in the model-building process. The variables are determined by correlation analysis to solve the factor selection problem. Data outliers may be detected by the K-NN algorithm to broaden applicability [25-27, 35, 36]. In addition, model quality is assessed by the leave-one-out algorithm [28-33]. Based on the big training dataset and the statistical learning methods, this paper proposed an efficient model that does not rely on time-consuming experiments. Finally, the adsorbed gas contents from nine reservoirs in China, Germany, and the U.S.A. were estimated as case studies.

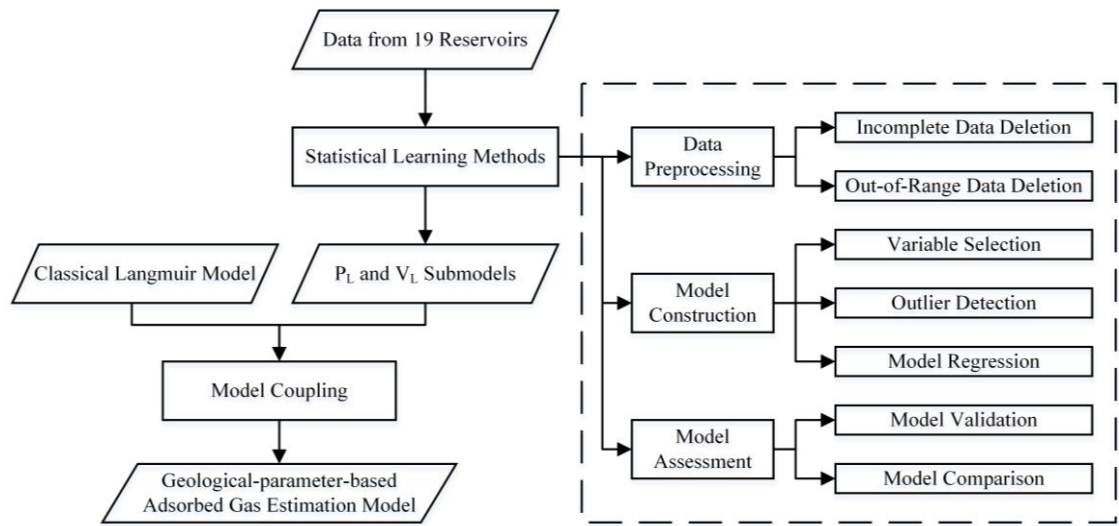

**Fig. 2.** Flow chart of the model-building process. The geological-parameter-based adsorbed gas estimation model is constructed by the coupling of classical Langmuir model, $P_L$ submodel, and $V_L$ submodel. The $P_L$ and $V_L$ submodels are built via statistical learning methods, which are explained in the dashed box.

## 2. Data description and preprocessing

The expansion of data size is necessary for developing an accurate estimation model, since it constitutes the foundation of the statistical learning process. Most existing models have no more than 10 data entries, which is not sufficient to construct a credible model. To overcome the data size restriction, raw data points are collected from 24 different studies, experiment reports, and databases in this study [23, 37, 39, 40, 44-46, 48-64]. These data are available in the online supplementary material of this paper. They are gathered from marine and terrestrial shales from various reservoirs around the world. **Table 1** shows the seven countries and 19 reservoirs used in this research. The variety of data sources makes the model suitable for general use.

**Table 1**

Data source information (detailed data are shown in the online supplement)

| Country | Reservoirs | Literature | Amount of data points |
| --- | --- | --- | --- |
| U.S.A. | Barnett Shale | [39, 48, 57] | 32 |
| | Haynesville Shale | [39] | 2 |
| | Marcellus Shale | [57-59] | 15 |
| | Utica Shale | [59] | 6 |
| | Woodford Shale | [46] | 3 |
| | Eagle Ford Shale | [39, 57] | 2 |
| Canada | Montney Shale | [57] | 2 |
| | Besa River Formation | [59] | 7 |
| | Colorado Group Formation | [59, 63] | 13 |
| | Exshaw Formation | [59, 64] | 3 |
| | Muskwa Formation | [59, 62] | 3 |
| | Duvernay Formation | [62] | 4 |
| China | Sichuan Basin | [23, 45, 48, 49, 51-53, 55, 56] | 136 |
| | Yangtze Platform | [40] | 21 |
| | Ordos Basin | [50, 61] | 18 |
| Germany | Posidonia Shale | [39, 60] | 31 |
| Sweden | Alum Shale | [39] | 20 |
| Netherlands | Carboniferous Shale | [39] | 9 |
| Brazil | Paraná Basin | [44] | 2 |

The raw data should be cleaned prior to use for model building. The cleaning process is explained in the online supplementary material. In this work, data preprocessing includes two straightforward steps. The first step is data cleaning, with the criterion that each data point should have both dependent variables and independent variables. According to this criterion, the $P_L$ data point should have $R_o$, TOC and T data, while the $V_L$ data point should have TOC and T data (Section 3.1 will introduce the independent variable selection method). In this step, incomplete data are deleted, and replicate data are integrated. The second step concerns the value range of geological parameters. It is reasonable to omit the out-of-range data. Previous work indicates that the TOC is always greater than 2% in promising shale gas reservoirs [13, 65]. To be conservative, 1% was taken as the floor level of TOC in this work. The $R_o$ of promising reservoirs should be lower than 4%, otherwise it is over-matured [66-68]. Moreover, the experimental data set exhibits the following properties: i) there is little data in which TOC is greater than 17% or T is greater than 90℃; ii) $P_L$ is usually greater than 1.5 MPa and less than 12 MPa; and iii) $V_L$ is the maximum adsorbed gas volume, and it has to be greater than 1 $m^3$/t for the sake of economic exploitation. Thus, the effective data point for $P_L$ should have a temperature below 90℃, $R_o$ less than 4%, TOC between 1% and 17%, and $P_L$ higher than 1.5 MPa and less than 12 MPa. After data processing, there are 101 data points remaining, as shown in the supplementary material online. Regarding $V_L$, the effective data points should have a temperature below 90℃, TOC between 1% and 17%, and $V_L$ higher than 1 $m^3$/t. The data points outside of this range have to be deleted. There are then 200 data points remaining, as shown in the supplementary material online. Finally, this research's database consists of 301 data points (101 $P_L$ data and 200 $V_L$ data), which is much larger than previous works' databases with fewer than 10 data points [44-47].

## 3. Model Construction

Regression models are built to identify the quantitative relationships between Langmuir and geological parameters. The model-building method consists of the following three parts: variable selection, outlier detection, and model regression.

*3.1 Variable selection*

In the Langmuir equation, $P_L$ and $V_L$ indicate the adsorption capacity of the shale formation. $V_L$ denotes the maximum adsorbed gas volume, and $P_L$ represents the pressure when the adsorbed gas volume is equal to half of $V_L$ [21]. In this work, the Langmuir parameters could be estimated by geological parameters due to the strong relationships between them. The geological variables are selected according to the data availability and their influences on $P_L$ and $V_L$. Recent studies have shown that the main influencing factors of $P_L$ are reservoir temperature (T), total organic carbon (TOC), and vitrinite reflectance ($R_o$, which reflects the thermal maturity). In addition, T, TOC, $R_o$, and porosity have effects on $V_L$. The qualitative relationships between Langmuir and geological parameters are shown in **Table 2**.

**Table 2**

Qualitative relationships between Langmuir and geological parameters based on literature values [23, 37-43]

|       | TOC      | $R_o$    | T        | Porosity |
|-------|----------|----------|----------|----------|
| $V_L$ | Positive | Negative | Negative | Positive |
| $P_L$ | Negative | Negative | Positive |          |

    Microscopically, the abundance of pores directly affects the adsorption capacity. Organic matter offers numerous pores, leading to large surface area and adsorption capacity [69]. This indicates that organic matter is not only hydrocarbon-generation matter, but also the carrier of adsorbed gas. So, $P_L$ decreases and $V_L$ increases with a higher TOC. Regarding thermal maturity, it reflects the thermal evolution degree of shale formations. Thus, within certain limits, a higher $R_o$ represents a higher hydrocarbon-generation potential [13, 66]. However, this does not guarantee low $P_L$ and high $V_L$ values for the following reasons. On the one hand, $R_o$ means vitrinite reflectance, which increases while organic carbon decomposes and becomes more mature [65]. A higher $R_o$ gives less organic carbon, and therefore reduces adsorbed gas potential. On the other hand, the organic carbon maturing process includes dehydrogenation and deoxygenation [70, 71], which will form pores in the microscopic structure and increase adsorbed gas potential [72, 73]. So, the influence of $R_o$ on adsorption ability should be determined by experimental results rather than speculation. In fact, adsorption experimental data indicate that $R_o$ is negatively correlated with both $P_L$ and $V_L$. Temperature is also an important factor for $P_L$ and $V_L$. According to Le Chatelier's principle, the increase of temperature inhibits the adsorption process, an exothermic process, which leads to lower adsorption capacity. Finally, the increase in temperature results in the increase of $P_L$ and decrease of $V_L$.

    Regarding the variable selection of the $P_L$ submodel, T, $R_o$, and TOC are chosen as the independent variables. For the sake of simplicity, the number of independent variables should be reduced by creating interaction terms and making the model three-dimensional. From the analysis above, we know that $R_o$ has a negative correlation with $P_L$, while there is a positive correlation between $P_L$ and T. Thus, according to the qualitative relationships between these independent variables, the interaction terms $\frac{T}{Ro}$ could be considered as a new independent variable to fit the $P_L$ submodel.

    Regarding the variable selection of the $V_L$ submodel, $R_o$ and porosity are eliminated based on data size, correlation coefficient between variables, and the cross-validation results. The data size is counted for each pair of variables. For instance, the data size of T vs. TOC means the number of data points which have effective $V_L$, T, and TOC simultaneously. The "effective data" here holds the same definition as that in Section 2. Correlation tests between different independent variables are then validated. The original dataset is given in the online supplementary Table C. The results are shown in **Table 3**.

**Table 3**

Data sizes and correlation coefficients between different independent variables

|  | Data Size | Correlation Coefficient (absolute value) |
|---|---|---|
| T vs. TOC | 200 | 0.26 |
| T vs. $R_o$ | 95 | 0.64 |
| T vs. Porosity | 67 | 0.32 |
| TOC vs. $R_o$ | 80 | 0.45 |
| TOC vs. Porosity | 88 | 0.57 |
| $R_o$ vs. Porosity | 34 | 0.39 |

As shown in **Table 3**, there is a relationship between $R_o$ and the other variables. The absolute correlation coefficient between T and $R_o$ is as high as 0.64, and that of TOC and $R_o$ is 0.45, which means that they are both moderately correlated [74]. In principle, $R_o$ increases while organic carbon decomposes, indicating a negative correlation between Ro and TOC. In addition, reservoir temperature affects the organic carbon maturing process, which is also measured by $R_o$. This means that the effect of TOC and T in this model may replace the one of $R_o$. Moreover, the data sizes of TOC vs. $R_o$ is only 80. The $V_L$ dataset only consists of 79 data points if the model takes all of the TOC, T, and Ro as variables. However, the data size has a 2.5-fold increase to 200 by cutting out $R_o$. In conclusion, on the one hand, $R_o$ is statistically and theoretically related to T and TOC. On the other hand, the availability of $R_o$ restricts the data size. So, it is reasonable to use T and TOC to substitute the effect of $R_o$ for the simplification of the model. According to the cross-validation results (Section 4.1 presents the process in detail), the elimination of $R_o$ actually increased the model accuracy. The average relative error decreased slightly from 25.75% of the model with $R_o$ to 23.76% without $R_o$. Furthermore, the half-width of the 90% confidence interval of relative errors dropped from 5.04% with $R_o$ to 2.33% without $R_o$, which indicates that the estimation accuracy of the model with $R_o$ has a higher fluctuation. The elimination of $R_o$ not only reduces the model dimensionality, but also expands the data size, which results in a more accurate estimation model.

Similarly, porosity is eliminated in the process. On the one hand, the size of the $V_L$ dataset can be expanded from 66 to 200 by leaving it out. On the other hand, the TOC and porosity are moderately correlated with the correlation coefficient of 0.57. This positive correlation has been observed in former studies [75]. Wang et al. determined that the porosity of organic matter in shale is higher than that of mineral matrix [76], which theoretically explicates the moderate correlation. Only one variable between TOC and porosity needs to be considered in the model. Since there are 200 data points that have T and TOC, and the correlation coefficient is only 0.26, porosity was substituted by TOC and T in the model. Furthermore, the cross-validation results show that the average relative error of the model with porosity (27.98%) is higher than the model without porosity (23.76%). In addition, the half-width of the 90% confidence interval of the model with porosity is 5.78%, which is more than two times larger than that of the model without porosity (2.33%). Thus, the $V_L$ submodel takes T and TOC as the independent variables.

*3.2 Outlier detection*

Outlier detection is essential because of the discrepancy of data sources and the uncertainty of geological parameters. Since geological variables have continuity, and the relationships between Langmuir parameters and geological variables are monotonous, the K-Nearest Neighbors algorithm (K-NN) is suitable for outlier detection [25-27, 35, 36]. The main idea of the K-NN algorithm is to classify the test data according to the k nearest training samples [77]. If the test data's property value does not match that of the training samples, the test data are considered to be an outlier and should be rejected. In this work, we define the weighted relative error ($R_i$) as the criterion of outlier detection, which means the relative difference between the i$^{th}$ test data's dependent variable value and the average value of the k nearby training samples. In this research, k is assigned to be 5 and the weighted relative error is defined with 1-norm, which is shown in **Eq. (2)**. The detailed formula derivation process is shown in **Appendix A**. The test data are regarded as an outlier if $R_i$ is greater than the given threshold. The denominator of $R_i$ is the smaller one of the test data's dependent value and the average value of training samples, while the numerator is the weighted sum of the difference between each training sample and test data. There is an inverse relationship between the weight and the distance to the neighbor, which means that the statistically nearer neighbors matter more than the statistically distant ones. The sum of the weights is equal to 1 for every test data.

$$R_i = \frac{\sum_{j=1}^{k} wR_{i,j} \times |V_{L,i} - V_{L,i,j}|}{\min\left(\frac{\sum_{j=1}^{k} V_{L,i,j}}{k}, V_{L,i}\right)} \tag{2}$$

where $V_{L,i}$ is the dependent variable value of the i$^{th}$ test data; $V_{L,i,j}$ represents the dependent variable value of the j$^{th}$ neighboring training data of the i$^{th}$ test data; $wR_{i,j}$ is the weight of the dependent value difference between the i$^{th}$ test data and the j$^{th}$ neighboring training data; and $wR_{i,j}$ is very significant, and this weight is calculated based on the "distance" between the test data and training data. Although the detailed calculation process is given in **Appendix A**, it is essential to introduce the "distance", since it is fundamental to determine the k nearest training samples and calculate the weight. First of all, this "distance" should not be the spatial distance, because the spatial relative location cannot precisely describe the similarity in adsorption ability between two different samples in the shale formation. The statistical distance which is based on geological parameters is applied to measuring the "distance" between different samples, in the form of Euclidean distance. Secondly, only the relative values of the variables in their distribution matter, and the absolute values are insignificant in the calculation of the statistical distance. Thus, the Euclidean distance should be adjusted, and it is necessary to perform normalization on these variables. As a result, the weighted Euclidean distance, which is a statistical distance, is computed to determine the k nearest neighbors and assign different weights to different samples.

There are 301 data points in the original dataset before the outlier detection, consisting of 101

$P_L$ data points and 200 $V_L$ data points. According to **Eq. (2)**, the weighted relative error (R) of each data point is calculated based on the K-NN algorithm. The results are shown in the online supplementary material Table B. If the R value of a data point is larger than 0.85, this data point is regarded as an outlier. Regarding the $P_L$ submodel, the weighted Euclidean distance is computed by the TOC, T, and $R_o$. After outlier detection, 10 data points are deleted, comprising 9.9% of the original data. Finally, 91 data points remain to fit the $P_L$ submodel. As for $V_L$, the weighted Euclidean distance is calculated based on T and TOC. According to R values, there are 16 data points deleted, taking up 8% of the original data. Finally, 184 data points remain to develop the $V_L$ submodel.

*3.3. Model regression*

The regression submodels were developed based on the processed data. The model construction process includes the following steps. First, according to previous qualitative studies [23, 37-43], the selected variables are combined to construct functions, which have clear physical meaning and succinct form. Second, mathematical adjustments are applied to variables to better describe the trend of these variables. Then, the $P_L$ and $V_L$ submodels are derived from the processed data, and the coefficients are identified by ordinary least square method (OLS) based on a normal equation [78]. Finally, the multivariate regression model between Langmuir and geological parameters is built based on the $P_L$ and $V_L$ submodels. This regression model can be used for adsorbed gas estimation.

Based on the cleaned data and qualitative relations, the simplified $P_L$ and $V_L$ submodels were constructed. A decay characteristic of Langmuir parameters could be observed in the data scatter plot. Thus, logarithmic treatment is applied to the dependent variables. The adjusted $P_L$ and $V_L$ submodels are shown in **Eq. (3)** and **(4)**, respectively. $TOC^*$ represents dimensionless total organic carbon and is equal to $\frac{TOC}{4\%}$. $T^*$ stands for $\frac{T}{48°C}$. It is the dimensionless adsorption temperature. $R_o^*$ means the dimensionless thermal maturity, which is equal to $\frac{Ro}{1.75\%}$. In this model, 4%, 48°C, and 1.75% are the average value of TOC, T, and $R_o$, respectively.

$$\ln P_L = a_P \cdot TOC^* + b_P \cdot \ln(\frac{T^*}{R_O^*}) + c_P \qquad (3)$$

$$\ln V_L = a_V \cdot TOC^* + b_V \cdot T^{*3} + c_V \qquad (4)$$

Using 91 data points as fitting data, the coefficients in the $P_L$ submodel were determined, which are $a_p$ = -0.136, $b_p$ = 0.715, and $c_p$ = 1.666. All of the variables are monotonously related to $P_L$ in the model. There is a positive correlation between $P_L$ and T, a negative correlation between $P_L$ and TOC, and a negative correlation between $P_L$ and $R_o$. These correlations match the qualitative analyses from previous work [23, 37-43]. **Fig. 3a** shows the distribution of $P_L$ data points and its comparison with the regression model. The blue points represent the data for regression, and the red stars are outliers detected by R value and the K-NN algorithm. Regarding the $V_L$ submodel, the coefficients were obtained from fitting the regression model, which are: $a_v$ = 0.421, $b_v$ = -0.067, and $c_v$ = 0.563. **Fig. 3b** shows the distribution of $V_L$ data points and the outliers. It compares all of the $V_L$ data points with the regression model. The regression model fits the experimental data well by visual comparison. As **Fig. 3b** shows, $V_L$ increases with decreasing T and increasing TOC. These results match the qualitative relationships mentioned in previous work [23, 37-43].

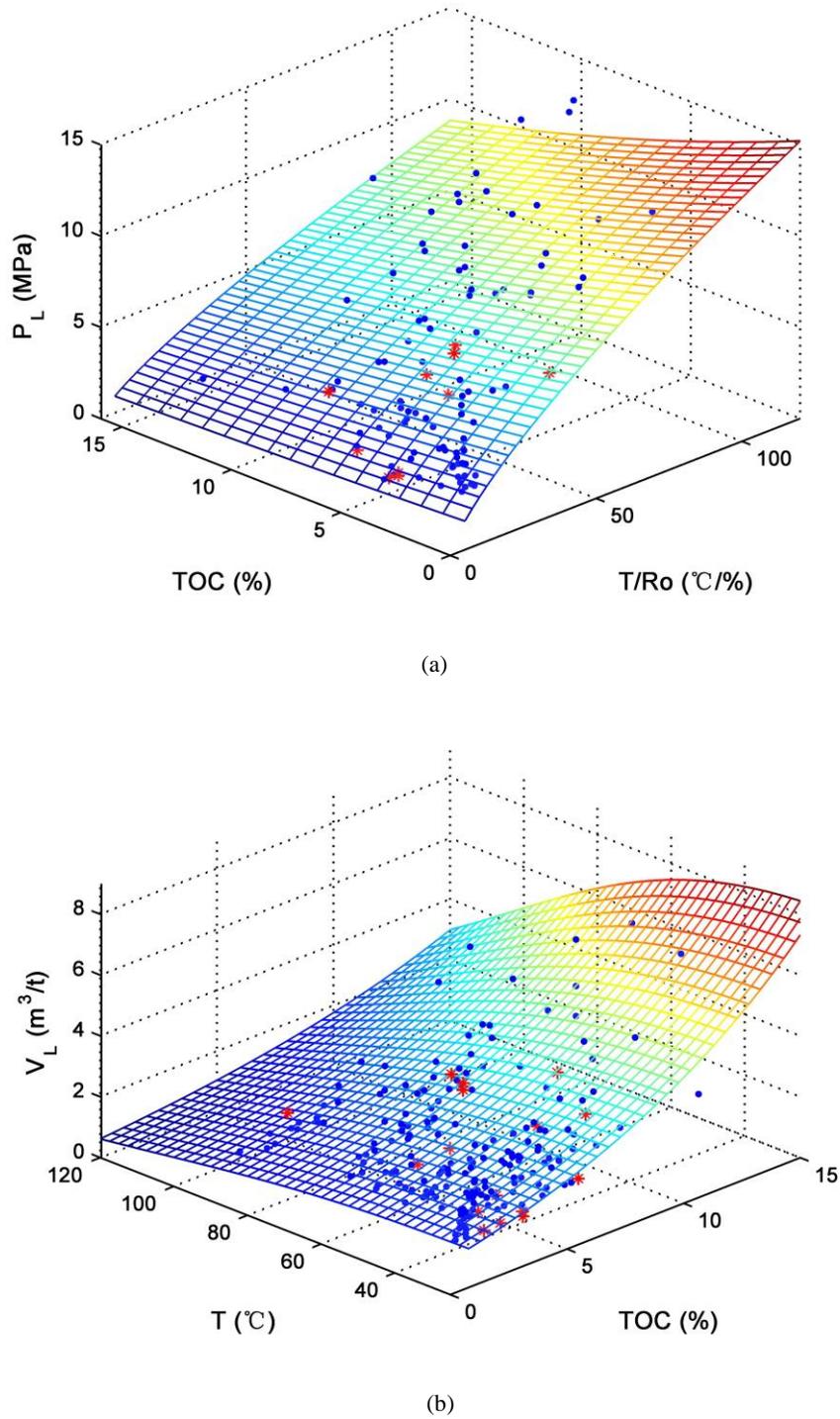

(a)

(b)

**Fig. 3.** Comparison of fitting data and outliers with the regression model: (a) $P_L$ data set and $P_L$ model; and (b) $V_L$

data set and $V_L$ model. The blue points describe the distribution of $P_L$ and $V_L$ data. The red stars represent the outliers in the model. The color grids show the regression model.

## 4. Model Assessment

*4.1. Model validation*

The visual comparison in **Fig. 3** is intuitive, but not precise. In order to obtain a reliable and stable model, cross-validation is necessary for the $P_L$ and $V_L$ submodels. The leave-one-out method [28-32] is used for cross-validation in this research. This method involves using one observation as the validation sample and taking the remaining data as the training set. The training set is utilized to fit the validation model, and the relative error is calculated by the validation sample.

When implementing the cross-validation, a vectorizing treatment can simplify the calculation. To vectorize the problem, three matrices are defined. Matrix X includes the original data points, of which each line describes an experimental data point, and each column corresponds to an independent variable. Matrix W is the coefficient matrix of the regression model. Matrix Y includes the values of the dependent variable of experimental data points. The $P_L$ submodel is taken as an example to introduce the vectorizing process. The $x_{in}$ (for i=1, 2, 3…m) is assumed to be 1. Thus, the $w_n$ represents the constant term in the model. The relationship of these matrices is shown in **Eq. (5)**.

$$\begin{bmatrix} x_{1,1} & \cdots & x_{1,n} \\ \vdots & \ddots & \vdots \\ x_{m,1} & \cdots & x_{m,n} \end{bmatrix} \times \begin{bmatrix} w_1 \\ \vdots \\ w_n \end{bmatrix} = \begin{bmatrix} P_{L,1} \\ \vdots \\ P_{L,m} \end{bmatrix} \quad (5)$$

The relative error between the validation sample and the training set is the indicator of the model quality. In the cross-validation process, all of the data points should be treated as a validation sample once. The i$^{th}$ data point is taken as an example to illustrate the leave-one-out process. The function to calculate the i$^{th}$ relative error is defined in **Eq. (6)**. $P_{L,i}$ is the experimental dependent variable of the i$^{th}$ data point, and $P'_{L,i}$ is the estimation of the dependent variable of the i$^{th}$ data point. This relative error is different with the weighted relative error (R) in **Eq. (2)**. The weighted relative error (R) is based on K-NN algorism and is only used in the outlier detection process. **Eq. (7)** presents the method to calculate $P'_{L,i}$, where $X_i$ is constructed with all of the data except the i$^{th}$ data point. The detailed formula derivation process is shown in **Appendix B**. The above process is repeated m times until all of the m samples are considered as validation data once. Finally, the relative errors of all samples are averaged to represent the accuracy of the $P_L$ and $V_L$ submodels. The confidential interval is calculated, as well.

$$Error_i = \frac{P_{L,i} - P'_{L,i}}{P_{L,i}} \times 100\% \tag{6}$$

$$P'_{L,i} = \begin{bmatrix} x_{i,1} & \cdots & x_{i,n} \end{bmatrix} \times \left(X_i^T X_i\right)^{-1} X_i^T \begin{bmatrix} P_{L,1} \\ \vdots \\ P_{L,i-1} \\ P_{L,i+1} \\ \vdots \\ P_{L,m} \end{bmatrix} \quad \text{where } X_i = \begin{bmatrix} x_{1,1} & \cdots & x_{1,n} \\ \vdots & \ddots & \vdots \\ x_{i-1,1} & \cdots & x_{i-1,1} \\ x_{i+1,1} & \cdots & x_{i+1,n} \\ \vdots & \ddots & \vdots \\ x_{m,1} & \cdots & x_{m,1} \end{bmatrix} \tag{7}$$

Regarding the $P_L$ submodel, all of the 91 data points are separately used as the validation data in the leave-one-out process. Finally, the average relative error with the 90% confidential interval is 27.67 ± 3.54 %. Considering the great variety of data sources, shale locations and shale types, this error is acceptable. The cross-validation result attests to the accuracy of the model. Furthermore, although the training data are different in every validation test, all of the fitting models are similar to the final regression model, which means that this model is robust and not sensitive to the training data. Based on the cross-validation results, **Fig. 4** shows the distribution of the experimental values, the estimated values, and the relative error between the experimental and estimated data. **Fig. 4a** presents the scatter of the experimental values versus their corresponding estimated values. The closer the distribution of points is to the 45° diagonal, the better the estimation. **Fig. 4b** is a normal Q-Q plot of the 91 average relative errors. This figure compares the probability distribution of the average relative errors with the normal distribution by plotting their quantiles against each other. The reference line represents the ideal condition when the relative errors obey the normal distribution exactly. It is clear from **Fig. 4** that the estimated values are similar to the experimental ones, and the relative errors are approximately normally distributed. Regarding the $V_L$ submodel, all of the 184 data points are used as the validation data. The relative errors are calculated just like for the $P_L$ submodel. Finally, the average relative error is 23.76%, with its 90% confidence interval being (21.43%, 26.09%). The confidence intervals are calculated based on a fixed-sample-size procedure [79]. **Fig. 5a** shows the scatter of the experimental values versus the estimated values, and **Fig. 5b** presents the Q-Q plot of the relative errors versus the normal distribution. As shown in **Fig. 4b** and **Fig. 5b**, there is a deviation between the expected normal values and average relative errors when the expected normal value increases beyond 40%. This deviation is resulted from the following two reasons. Firstly, the acceptance region of Q-Q plot is not parallel to the reference line. The acceptance region becomes wider when the quantile gets closer to 0 or 1 (corresponding to the bottom left and upper right of the Q-Q plot) [80, 81]. Secondly, the model has less data points with large positive error compared with the normal distribution, which results in the phenomena that expected values are above the reference line on the upper right. This actually means that the relative error in the model is apt to be smaller than that in the normal distribution when the error is above 40%. In addition, the 40% relative error corresponds to the 91% quantile in standard normal distribution. Thus, the deviation in the Q-Q plot actually only affects a small proportion of the data, and this means that there are fewer positive errors in the model than the normal distribution.

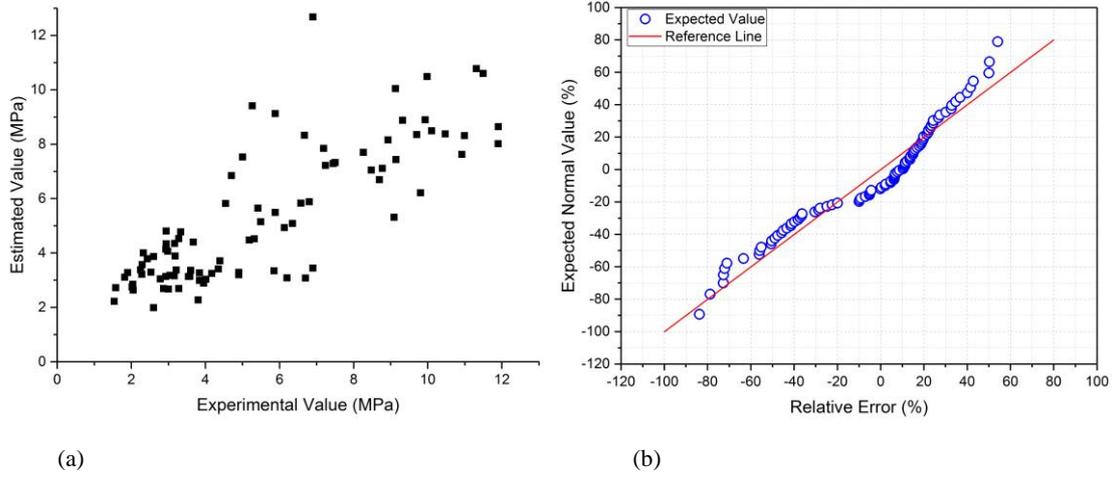

(a)                                                 (b)

**Fig. 4.** Results of the $P_L$ submodel's cross-validation: (a) the scatter of the experimental value versus its corresponding estimated value; and (b) the Q-Q plot of the distribution of the relative errors versus the normal distribution.

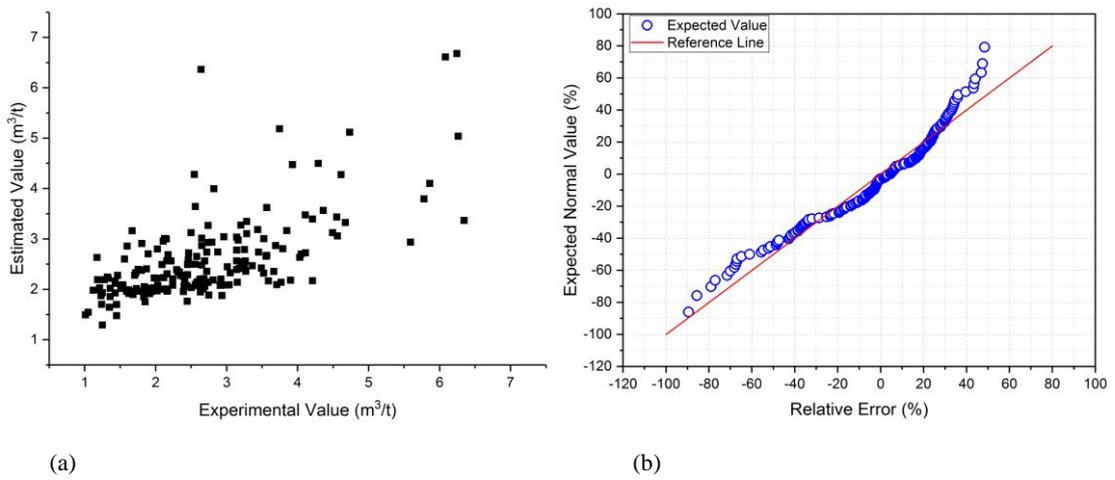

(a)                                                 (b)

**Fig. 5.** Results of the $V_L$ submodel's cross-validation: (a) the scatter of the experimental value versus its corresponding estimated value; and (b) the Q-Q plot of the distribution of the relative errors versus the normal distribution.

*4.2. Model comparison*

    Some previous work has already proposed a couple of $P_L$ and $V_L$ submodels based on geological parameters, as listed in **Table 4** [23, 44-46, 82]. However, the inadequate training data and the lack of independent variables affect the estimation accuracy and applicability of these models. Most of those studies' training sets have less than 10 data points due to the time-consuming adsorption experiments and the expensive coring process. Previous models and the model developed in this research were compared by calculating the average relative errors. The process is as follows. First, the training data are randomly extracted from the entire dataset to fit the regression model. The proportion of the training data is 80% and 90% for the $P_L$ and $V_L$ submodels, respectively. Then, the other 20% (for the $P_L$ submodel) and 10% (for the $V_L$

submodel) of the entire data are used as the test data for the average relative error calculation to validate the model. Since the economically attractive shale gas reservoirs always have relatively high TOC and high $R_o$, it is also essential to compare the different models under these conditions. The extraction and validation process are performed 14 times for the $P_L$ submodels and 11 times for the $V_L$ submodels. The first five times for each model are of the entire dataset extraction, which means that the test data are extracted randomly from all of the 91 data points for the $P_L$ submodels and from the 184 data points for the $V_L$ submodels. The next three times are for the high temperature scenario, in which the test data are stochastically sampled from the data whose temperature is higher than 65℃. There are also three comparisons belonging to the high TOC scenario, in which the test data are taken from the data whose TOC are higher than 5%. The last three comparisons of the $P_L$ submodels are for the high $R_o$ scenario, in which the test data are extracted from the data with $R_o$ higher than 2%. Finally, the fitting precisions of these models are compared based on the cross-validation results.

**Table 4**

Different $P_L$ and $V_L$ submodels from previous studies

| Literature | Model |
| --- | --- |
| [46] | $\ln \dfrac{1}{P_L} = \dfrac{a}{T} + c$ |
| [82] | $P_L = a \times TOC^{-b}$ |
| [82] | $V_L = a \times TOC^{b}$ |
| [23, 44-46] | $V_L = a \times TOC + b$ |

Concerning the $P_L$ submodels, 18 data points are extracted from the $P_L$ dataset as the test data. The validation process is repeated 14 times. The calculation results are shown in **Table 5**. The overall average relative error of the proposed $P_L$ submodel is 25.13%, which is much less than the 35.90% and 40.04% from the existing models. In the high temperature scenario, the comparisons between the existing and new models are independently conducted three times. Since the shale gas reservoirs are generally deep, many reservoirs' temperatures are higher than 65℃ [83-87]. The average relative errors of the existing models are 47.77% and 42.56%, and that of the new model is 26.25% in the high temperature scenario, indicating that the new model possesses an advantage over the previous ones. Considering real-world applications, the high TOC and high $R_o$ scenarios should be examined as well, because commercially attractive shale gas reservoirs always have high TOC and $R_o$. The advantage of the new model is obvious in these scenarios. The average relative errors decrease from 46.01% and 48.20% (the existing models) to 27.52% (the new model) under the high TOC condition. In the high $R_o$ scenario, the errors drop from 60.52% and 71.33% (the existing models) to 33.60% (the new model). It is clear that the new model's average relative errors are much less than the existing models from the literature, not only under the overall condition but also in the circumstances of high temperature, high TOC, and high $R_o$. Considering real-world situations, the new $P_L$ submodel is more useful and accurate.

**Table 5**

Relative error comparison of the $P_L$ submodels in different scenarios

| Test Number | Relative Error (%) | | |
|---|---|---|---|
| | $\ln\dfrac{1}{P_L}=\dfrac{a}{T}+c$ | $P_L=a\times TOC^{-b}$ | $\ln P_L=a\times TOC^{*}+b\times\ln\dfrac{T^{*}}{Ro^{*}}+c$ |
| Test 1 | 40.78 | 35.21 | 23.42 |
| Test 2 | 35.20 | 45.14 | 27.56 |
| Test 3 | 23.74 | 32.61 | 22.15 |
| Test 4 | 48.88 | 52.02 | 30.51 |
| Test 5 | 30.89 | 35.19 | 21.99 |
| **Average** | **35.90** | **40.04** | **25.13** |
| HighT1 | 50.15 | 44.54 | 24.84 |
| HighT2 | 52.54 | 43.11 | 28.01 |
| HighT3 | 40.63 | 40.02 | 25.89 |
| **Average** | **47.77** | **42.56** | **26.25** |
| HighTOC1 | 30.51 | 33.26 | 26.20 |
| HighTOC2 | 46.78 | 40.52 | 26.57 |
| HighTOC3 | 60.74 | 70.82 | 29.79 |
| **Average** | **46.01** | **48.20** | **27.52** |
| HighR$_o$1 | 55.39 | 79.44 | 32.44 |
| HighR$_o$2 | 67.91 | 70.30 | 39.78 |
| HighR$_o$3 | 58.26 | 64.26 | 28.57 |
| **Average** | **60.52** | **71.33** | **33.60** |

For the $V_L$ submodel, the new model is compared with the existing models in different scenarios, as well. **Table 6** shows the validation results. According to **Table 6**, the average relative error of the new model is 23.87%, which is less than 25.42% and 25.79% for the existing models, among the five entire dataset extraction tests. This means that the new model is more accurate, in general. Furthermore, regarding the high temperature scenarios, the average relative errors are 38.45% and 39.59% for the existing models, and 23.34% for the new model. This means that the introduction of the temperature term to the model reduces the relative error by more than 15%. Furthermore, the three tests with high TOC indicate that the new $V_L$ submodel has a similar performance to the existing models. As such, the new $P_L$ and $V_L$ submodels are statistically better than the existing models, especially under real-world conditions. Therefore, the proposed $P_L$ and $V_L$ submodels could be applied to Langmuir parameters estimation under real-world conditions.

**Table 6**

Relative error comparison of the $V_L$ submodels in different scenarios

| Test Number | Relative Error (%) | | |
|---|---|---|---|
| | $V_L=a\times TOC^{b}$ | $V_L=a\times TOC+b$ | $\ln V_L=a\times TOC^{*}+b\times T^{*3}+c$ |
| Test 1 | 23.34 | 21.24 | 21.27 |
| Test 2 | 29.32 | 30.86 | 23.06 |
| Test 3 | 26.07 | 27.52 | 25.06 |
| Test 4 | 24.68 | 26.36 | 25.67 |
| Test 5 | 23.68 | 22.98 | 24.30 |
| **Average** | **25.42** | **25.79** | **23.87** |
| HighT1 | 35.16 | 38.11 | 20.49 |
| HighT2 | 45.82 | 45.56 | 27.27 |
| HighT3 | 34.39 | 35.10 | 22.27 |
| **Average** | **38.45** | **39.59** | **23.34** |
| HighTOC1 | 23.42 | 24.87 | 22.44 |
| HighTOC2 | 25.43 | 25.17 | 25.46 |
| HighTOC3 | 23.08 | 28.12 | 24.42 |
| **Average** | **23.98** | **26.05** | **24.11** |

## 5. Geological-parameter-based estimation model

Since the Langmuir model is the most commonly used empirical model to estimate adsorbed gas content, we propose a new model for estimating the adsorbed gas content on the basis of the $P_L$ and $V_L$ submodels presented in the above. By plugging **Eq. (3)** and **(4)** into the Langmuir model (**Eq. (1)**), the $P_L$ and $V_L$ terms can be substituted by geological parameters, such as reservoir temperature, TOC, and $R_o$. All of these geological parameters can be determined without adsorption experiments. By replacing the Langmuir parameters with the geological parameters, the classical Langmuir model can be transformed to the new geological-parameter-based estimation model in **Eq. (8)**. This model only depends on the directly measurable geological parameters. The adsorbed gas content can be easily estimated by the reservoir depth h, TOC, $R_o$, reservoir pressure, and temperature. The regression coefficients, $a_P$, $b_P$, $c_P$, $a_V$, $b_V$, and $c_V$ are from the $P_L$ and $V_L$ submodels. They are constants when the training database is fixed.

$$V = \frac{V_L}{1+\frac{P_L}{P}} = \frac{\exp(a_V \cdot TOC^*) \cdot \exp(b_V \cdot T^{*3}) \cdot \exp(c_V)}{1+\frac{1}{P} \cdot [\exp(a_P \cdot TOC^*) \cdot (\frac{T^*}{R_O^*})^{b_P} \cdot \exp(c_P)]} \qquad (8)$$

Although the reservoir temperature and pressure in **Eq. (8)** can be measured directly, they can also be estimated when it is difficult to determine them. The reservoir temperature can be calculated by the reservoir depth (h), the surface temperature and the temperature gradient (gradT), as shown in **Eq. (9)**. In addition, the reservoir pressure can be determined by the reservoir depth and the pressure coefficient (α) as **Eq. (10)**. $\vec{n}$ is a normal unit vector, which is dimensionless. The reservoir isothermal surfaces are assumed to be horizontal, and thus the direction of $\vec{n}$ is vertically downward, which is the same as the reservoir depth's direction. $T_s$ represents the surface temperature, ℃. $P_h$ means the hydrostatic pressure, or the pressure exerted by the column of water above the formation, MPa. It can be estimated by the water density and the reservoir depth. α is the pressure coefficient, which is dimensionless. $\rho_w$ represents water density equal to 1, t/m³. g is the local acceleration of gravity equal to 9.8, N/kg.

$$T = h \cdot gradT + T_s = h \cdot \frac{\partial T}{\partial n} \cdot \vec{n} + T_s \tag{9}$$

$$P = \alpha \cdot P_h = \alpha \cdot \rho_w \cdot g \cdot h = 9.8 \cdot \alpha \cdot h \tag{10}$$

## 6. Case study: Adsorbed gas content estimation of nine shale gas reservoirs

To estimate the adsorbed gas content in different reservoirs, the values of the geological parameters in different reservoirs should be determined accordingly. In this study, the following reservoirs in China, Germany, and the U.S.A. are examined: the Sichuan Basin, the Yangtze Platform, the Songliao Basin, the Ordos Basin, the Tarim Basin, the Northern Jiangsu Basin, the Marcellus Shale, the Barnett Shale, and the Posidonia Shale. The Marcellus Shale and the Barnett Shale are from the U.S.A., and the Posidonia Shale is from Germany. The remaining six reservoirs are in China.

The data of TOC and $R_o$ were collected from previous work, which are shown in **Table 7**. **Fig. 6** presents the distribution of TOC and $R_o$ data for different reservoirs. It is clear from **Fig. 6a** that both the mean and median of TOC of all basins are higher than 2%, which is the commercial development threshold [13, 65]. **Fig. 6b** shows that, except for the Northern Jiangsu Basin, the Songliao Basin, the Barnett Shale, and the Posidonia Shale, the mean and median of other reservoirs' $R_o$ are higher than 1.3%. The $R_o$ values indicate that these reservoirs are relatively mature and have passed the oil window [88]. Regarding the Yangtze Platform, the Sichuan Basin, and the Marcellus Shale, their $R_o$ values exceed 2%, which indicates that they are already in a dry gas window [13, 70, 89, 90]. The $R_o$ distribution implies that the organic matter of these shale gas reservoirs is highly matured, and this explains why these shale reservoirs mainly produce shale gas, but not shale oil. Concerning the reservoir temperature, the Global Heat Flow Database provided by the International Heat Flow Commission [91] contains temperature gradient data from all around the world. The temperature gradient data within China, Germany, and the U.S.A. have 667, 254, and 4249 data points, respectively. The inverse distance weighting (IDW) interpolation method [92-94] is used to calculate reservoir temperature since temperature is a continuous variable underground. This interpolation method assumes that the value of an unsampled point is the weighted average of its neighbor samples. In addition, the weight is inversely related to the distance between the unsampled point and its neighbor. The temperature distribution of each reservoir is calculated in **Appendix C**, and the average temperature is used as the reservoir temperature in the following estimation.

**Table 7**
TOC and $R_o$ data sizes for different shale reservoirs [48, 95-113]

| Reservoir | Sichuan Basin | Yangtze Platform | Songliao Basin | Ordos Basin | Tarim Basin | Northern Jiangsu Basin | Marcellus Shale | Barnett Shale | Posidonia Shale |
|---|---|---|---|---|---|---|---|---|---|
| Amount of TOC data | 39 | 31 | 6 | 8 | 12 | 10 | 12 | 9 | 18 |
| Amount of $R_o$ data | 27 | 13 | 5 | 8 | 12 | 10 | 3 | 8 | 18 |

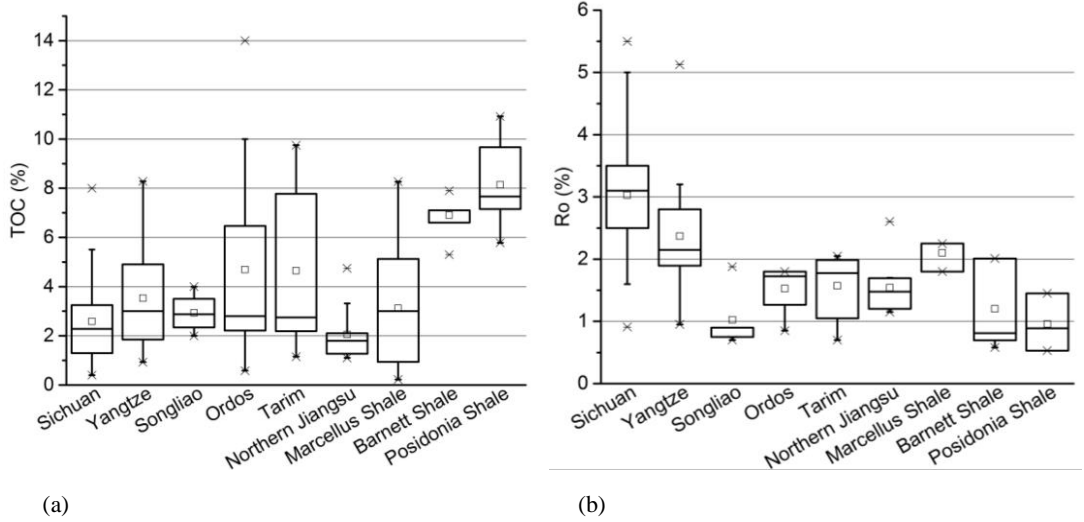

(a)                                                     (b)

**Fig. 6.** Distribution of the TOC and $R_o$ data for the Sichuan Basin, the Yangtze Platform, the Songliao Basin, the Ordos Basin, the Tarim Basin, the Northern Jiangsu Basin, the Marcellus Shale, the Barnett Shale, and the Posidonia Shale: (a) the TOC box plot of different reservoirs; and (b) the $R_o$ box plot of different reservoirs. The asterisks represent outliers that are out of the 1.5 IQR (interquartile range). The blank squares correspond to the means of TOC and $R_o$ in different reservoirs.

The geological parameters are evaluated separately, and the results are shown in **Table 8**. To calculate the reservoir pressure, each reservoir pressure coefficient is assumed to be 1. The surface temperature is assumed to be 20 ℃ when calculating the reservoir temperature in this study. Finally, the adsorbed gas contents are estimated by plugging the geological parameters into the geological-parameter-based estimation model (**Eq. (8)**). The results are shown in the last column of **Table 8** and **Fig. 7**. The observations of adsorbed gas content in the Sichuan Basin, the Ordos Basin, and the Marcellus Shale are shown in **Fig. 7**, as well. Owing to the high cost of adsorption experiments, there is a lack of experimental data about adsorbed gas contents in most reservoirs. However, the Sichuan Basin, the Ordos Basin, and the Marcellus Shale are relatively developed, and some researchers have estimated the adsorbed gas contents. Thus, their data can serve as a reference used to compare the estimation results in this study. According to previous studies, the adsorbed gas content in the Sichuan Basin ranges from 1.12 m³/t to 1.74 m³/t, with the average of 1.28 m³/t [114]. The estimation result in this work is 1.31 m³/t, which is consistent with previous work. Regarding the Ordos Basin, researchers have evaluated the adsorbed gas content in Chang 7 Member and Chang 9 Member of the Yanchang Formation, which are 1.17 - 3.68 m³/t, with the mean of 1.67 m³/t and 1.19 - 2.62 m³/t with the average of 1.32 m³/t, respectively [115]. These results confirmed the estimation in this research, in which the adsorbed gas content is 1.47 m³/t in the Ordos Basin. As for the Marcellus Shale, researchers have shown in a previous study that the adsorbed gas content ranges from 0.85 m³/t to 1.4 m³/t [101]. The estimation of our model is 1.24 m³/t, which confirms with the observation in the previous study. It is thus shown that the experimental results from previous studies validated the estimation in the Sichuan Basin, the Ordos Basin, and the Marcellus Shale from this research. Moreover, the estimation procedure is totally independent of adsorption experiments in this study, which reduces the estimation cost and increases efficiency.

**Table 8**

Geological parameters and the estimated adsorbed gas contents in China shale gas reservoirs

| Reservoir | Depth (m) | TOC (%) | $R_o$ (%) | Reservoir Temperature (℃) | Reservoir Pressure (MPa) | Adsorbed Gas Content (m³/t) |
|---|---|---|---|---|---|---|
| Sichuan Basin | 3230 | 2.58 | 3.03 | 86.98 | 31.65 | 1.34 |
| Yangtze Platform | 1737 | 3.53 | 2.37 | 57.48 | 17.02 | 1.81 |
| Songliao Basin | 1731 | 2.93 | 1.03 | 90.46 | 16.96 | 0.92 |
| Ordos Basin | 2730 | 4.69 | 1.53 | 86.93 | 26.75 | 1.51 |
| Tarim Basin | 4023 | 4.65 | 1.57 | 95.99 | 39.43 | 1.39 |
| Northern Jiangsu Basin | 2872 | 2.05 | 1.54 | 106.19 | 28.15 | 0.79 |
| Marcellus Shale | 2057 | 3.12 | 2.10 | 87.26 | 20.16 | 1.24 |
| Barnett Shale | 2286 | 6.90 | 1.20 | 83.23 | 22.40 | 1.88 |
| Posidonia Shale | 53 | 8.14 | 0.96 | 22.66 | 0.52 | 0.52 |

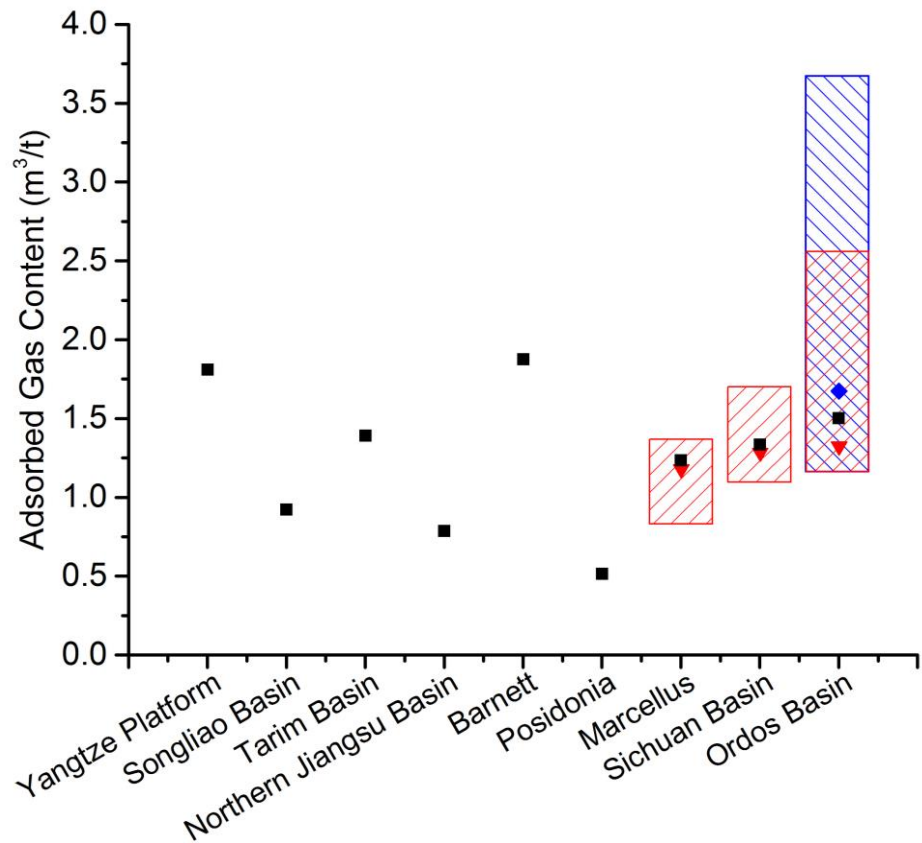

**Fig. 7.** Comparison of adsorbed gas content estimations and the observations in nine reservoirs. The black squares indicate the estimation values in nine reservoirs. The floating chart shows the distributions of adsorbed gas content

observations in the Marcellus Shale, the Sichuan Basin, and the Ordos Basin. Regarding the Ordos Basin, the red column in the floating chart describes the observations in the Chang 9 Member, and the blue column represents the Chang 7 Member.

**7. Discussion and Conclusion**

In this study, statistical learning methods were utilized to construct an adsorbed gas content estimation model based on geological parameters. Statistical learning is effective in finding a predictive function, and it can elucidate hidden patterns and unknown correlations between shale gas content and other variables. In the model construction process, 301 data points are collected from different studies, experiment reports, and databases. The outliers are detected by the K-NN algorithm, and the model performance is validated by the leave-one-out cross-validation. Moreover, the usage of geological parameters makes it possible to estimate adsorbed gas content without conducting time-consuming adsorption experiments and an expensive coring process. Thus, the geological-parameter-based estimation model is efficient with a relatively low cost.

Regarding variable selection of the estimation model, more variables may be considered in the model, as long as there are enough data to avoid overfitting. Considering the data availability, we only take T, TOC, $R_o$, and porosity as candidate variables in this paper. These four variables are also the most commonly used variables in the previous studies. Among the other variables, the clay content is important since the adsorption capacities of clay and kerogen are different. Lu et al. proposed a Bi-Langmuir model that can account for gas adsorption on both clay minerals and kerogen [116, 117]. Although our model can be extended to consider clay content without any change in the regression model, the clay content data are not currently widely available.

It should be mentioned that this model is not a complete substitution for adsorption experiments. However, this model is a good choice for the overall approximate estimation of gas content since it can reduce the estimation cost and increase estimation efficiency. This model also makes it possible to estimate the adsorbed gas content at a large scale, such as a whole reservoir. This model can also be applied to an approximate estimation for a single well at a micro scale when no site-specific measurements have been made. When the exact adsorbed gas content of a sample is expected, adsorption experiments still constitute the optimal choice with the availability of adequate equipment and sufficient time.

The geological-parameter-based estimation model is constructed by substituting the $P_L$ and $V_L$ terms in the classical Langmuir model by two submodels. Compared with the existing $P_L$ and $V_L$ submodels, the new submodels proposed in this research possess an advantage under real-world conditions. Regarding the $P_L$ submodel, the estimation errors decrease to nearly half of that of the existing models in the high temperature scenario, high TOC scenario, and high $R_o$ scenario. The overall average relative error from cross-validation is 27.67 ± 3.54 % with a 90% confidential interval. As for the $V_L$ submodel, the estimation relative error decreases apparently in the high temperature scenario, and is similar to the existing models in the high TOC scenario. The overall average relative error is 23.76%, with its 90% confidence interval being (21.43%, 26.09%). Because most promising reservoirs have high TOC and high Ro, the new submodels perform better under real-world conditions.

Finally, a geological-parameter-based estimation model is developed to estimate adsorbed gas content in shale gas reservoirs, and it is applied to the Sichuan Basin, the Yangtze Platform, the

Songliao Basin, the Ordos Basin, the Tarim Basin, the Northern Jiangsu Basin, the Marcellus Shale, the Barnett Shale, and the Posidonia Shale as case studies. A major advantage of this model is that the whole estimation process is totally independent of time-consuming adsorption experiments and the expensive coring process. As shale gas developments are still in the initial phase in many countries, there are no specific measurements available at most shale gas reservoirs. The model presented in this research provides a reasonably accurate approach to estimate the adsorbed gas contents in different reservoirs around the world with easily-obtained geological parameters.

**Acknowledgements**

This work is partially funded by the National Natural Science Foundation of China (Grant No. U1262204, 51520105005, and U1663208) and the National Key Technology R&D Program of China (Grant No. 2012BAC24B02). The data used in the estimation model is available in the online supplementary material of this paper.

**Appendix. A: Detailed formula derivation process for outlier detection**

The outlier detection process is described as follows. First, a distance should be calculated to determine k nearest training samples. As discussed in the body of the text, this "distance" should not be the spatial distance since it is not suitable for measuring the similarity in adsorption ability between two different samples. In addition, it is necessary to perform normalization on the variables due to the different scales of these variables. Thus, a weighted Euclidean distance, which is a statistical distance based on geological parameters, is used to determine the k nearest neighbors and to assign different weights to different samples. The distance is defined in **Eq. (A.1)**.

$$D_{i,j} = \sqrt{\sum_{l=1}^{n} \left[ w_l^2 \times (x_{l,i} - x_{l,j})^2 \right]} \tag{A.1}$$

where $D_{i,j}$ means the statistical distance between the i$^{th}$ sample and j$^{th}$ sample; $x_{l,i}$ represents the i$^{th}$ sample of the l$^{th}$ variable; and $w_l$ represents the weight of the l$^{th}$ variable. This weight is used for normalization. The most common normalization method is called min-max normalization which performs a linear transformation on the original data [118]. Usually, the new normalized value equals the difference between two points divided by the difference between the maximum and minimum value. This method maps the difference between two points to a value in the range from 0 to 1. However, the normalization process in **Eq. (A.1)** occurs prior to the outlier detection, which means that the data used might be an outlier. If the maximum or minimum values end up to be an outlier, the min-max normalization will be greatly affected, because the denominator in the weight will be abnormally large. To avoid this problem, we replace the difference between maximum and minimum values with the interquartile range (IQR) in this paper. The difference between maximum and minimum values is sensitive to the abnormal values, but the IQR is not.

Finally, $w_l$ is stated as **Eq. (A.2)**. This weight is related to the distribution of variables, and it is essential for the normalization.

$$w_l = \frac{10}{Q_{3,l} - Q_{1,l}} \tag{A.2}$$

where $Q_{1,l}$ and $Q_{3,l}$ represent the upper and lower quartile of the $l^{th}$ variable, respectively. Using this weighted distance, the impact of different variables on the statistical distance could be evaluated.

The outliers are then detected from the relative difference between the test data and the average value of k nearby training samples. In this research, k is assigned to be 5. The relative difference is defined with 1-norm in **Eq. (2)**, and the weight is shown in **Eq. (A.3)**. The test data are regarded as an outlier if $R_i$ is greater than the given threshold.

$$wR_{i,j} = \frac{\sum_{l=1}^{k} D_{i,l} - D_{i,j}}{(k-1) \times \sum_{l=1}^{k} D_{i,l}} \tag{A.3}$$

where $wR_{i,j}$ is the weight of the dependent value difference between the $i^{th}$ test data and the $j^{th}$ neighboring training data; and $\sum_{l=1}^{k} D_{i,l}$ is the sum of the statistical distances between the $i^{th}$ test data and all of its k neighboring training data.

Regarding the $P_L$ submodel, the weighted statistical distance function is defined as **Eq. (A.4)**. The relative error (R) is calculated according to **Eq. (2)**. The data points whose R values are greater than 0.85 are removed as outliers. Concerning $V_L$, since the regression model of $V_L$ uses T and TOC as the independent variables, the distance function is defined as **Eq. (A.5)**. The data points whose R value are greater than 0.85 are removed.

$$D_{i,j} = \sqrt{w_T^2 \times (T_i - T_j)^2 + w_{TOC}^2 \times (TOC_i - TOC_j)^2 + w_{Ro}^2 \times (Ro_i - Ro_j)^2} \tag{A.4}$$

$$D_{i,j} = \sqrt{w_T^2 \times (T_i - T_j)^2 + w_{TOC}^2 \times (TOC_i - TOC_j)^2} \tag{A.5}$$

**Appendix. B: Detailed formula derivation process for the leave-one-out method**

The $i^{th}$ data point is taken as a validation sample to illustrate the process of leave-one-out cross-validation. First, matrix X and Y are constructed with all data except the $i^{th}$ data point (**Eq. (B.1)**), and then W is calculated according to the normal equation (**Eq. (B.2)** and **(B.3)**) [78].

$$\begin{bmatrix} x_{1,1} & \cdots & x_{1,n} \\ \vdots & \ddots & \vdots \\ x_{i-1,1} & \cdots & x_{i-1,1} \\ x_{i+1,1} & \cdots & x_{i+1,n} \\ \vdots & \ddots & \vdots \\ x_{m,1} & \cdots & x_{m,1} \end{bmatrix} \times \begin{bmatrix} w_1^i \\ \vdots \\ w_n^i \end{bmatrix} = \begin{bmatrix} P_{L,1} \\ \vdots \\ P_{L,i-1} \\ P_{L,i+1} \\ \vdots \\ P_{L,m} \end{bmatrix} \tag{B.1}$$

$$X_i^T X_i \begin{bmatrix} w_1^i \\ \vdots \\ w_n^i \end{bmatrix} = X_i^T Y_i \tag{B.2}$$

$$\begin{bmatrix} w_1^i \\ \vdots \\ w_n^i \end{bmatrix} = \left( X_i^T X_i \right)^{-1} X_i^T Y_i \tag{B.3}$$

To estimate the dependent variable of the i<sup>th</sup> data point, the independent variables of the i<sup>th</sup> data point are multiplied by the matrix W (**Eq. (B.4)**). The function to calculate the relative error is defined as **Eq. (B.5)**. The dependent variable of the i<sup>th</sup> data point is expressed as $P'_{L,i}$. $P_{L,i}$ represents the experimental value of the i<sup>th</sup> data point's dependent variable.

$$\left[ P'_{L,i} \right] = \begin{bmatrix} x_{i,1} & \cdots & x_{i,n} \end{bmatrix} \times \begin{bmatrix} w_1^i \\ \vdots \\ w_n^i \end{bmatrix} \tag{B.4}$$

$$Error_i = \frac{P_{L,i} - P'_{L,i}}{P_{L,i}} \times 100\% \tag{B.5}$$

The above steps are repeated m times until all of the samples are considered as validation data once. Finally, the relative errors of all samples are averaged to represent the accuracy of the $P_L$ and $V_L$ submodels.

**Appendix. C: Temperature distribution of six Chinese reservoirs**

According to the Global Heat Flow Database [91], the scatter of gradT versus the average depth of the measuring section is shown in **Fig. C.1.** It is obvious that the gradT whose average depth is lower than 500 m fluctuates greatly. This means that the gradT with low depth is easily affected by the variation of the surface temperature. It is reasonable to ignore those data within 500 m, since most of the shale gas reservoirs are much deeper than 500 m in China (99.82% of the risked GIP is deeper than 500 m, and 40.46% is deeper than 3000 m [14]). Finally, there are 348 gradT data remaining to calculate reservoir temperature.

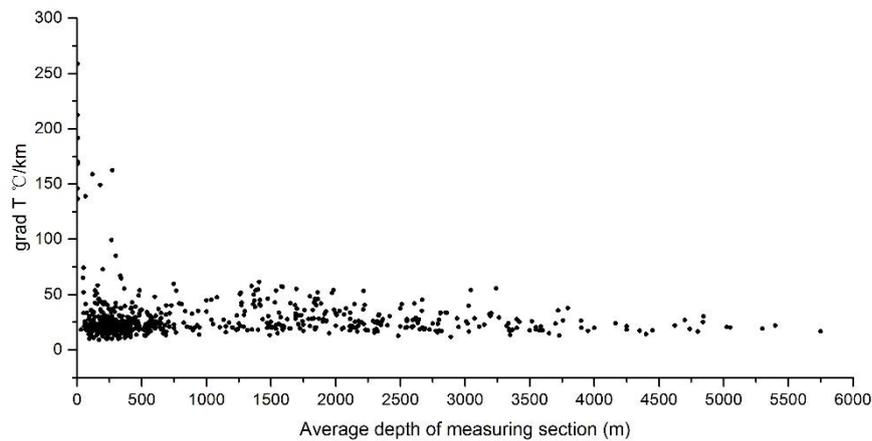

**Fig. C.1.** Scatter of the gradT versus average depth of the measuring section in China from the Global Heat Flow Database. The gradT whose average depth is lower than 500 m fluctuates greatly.

The reservoir temperature T is the sum of the surface temperature and the additional temperature calculated by gradT and depth. The inverse distance weighting (IDW) interpolation method [92-94] is applied to calculate the additional temperature. Because different shale gas reservoirs have different depths, the nephogram on a certain depth does not accord with the real situation in the reservoirs. Each reservoir's temperature should be calculated only based on this reservoir's average depth and its corresponding gradT. The result is shown in **Fig. C.2**. In this temperature nephogram, the IDW interpolation is applied within each reservoir totally independently from the others.

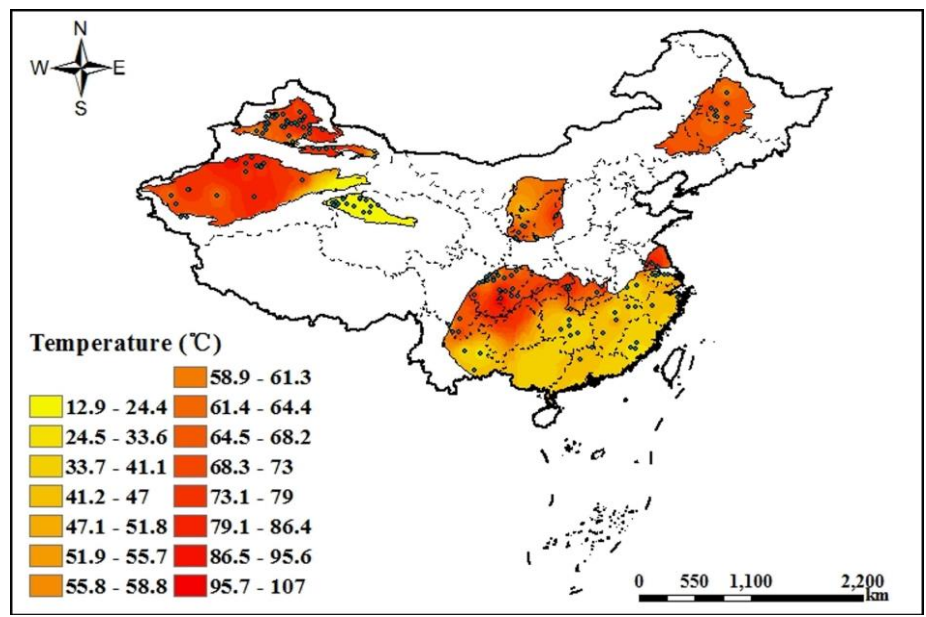

**Fig. C.2.** Additional temperature nephogram of China based on the real depth of different reservoirs.

# References

1.  Aguilera, R.F., *The role of natural gas in a low carbon Asia Pacific.* Applied Energy, 2014. **113**:

Geoscience, 2011. **22**(6): p. 1100-1108.